\documentclass[sigconf]{acmart}
\AtBeginDocument{%
  }
\setcopyright{acmlicensed}
\copyrightyear{2026}
\acmYear{2026}
\setcopyright{cc}
\setcctype{by}
\acmConference[IH '26]{Interactive Health Conference}{July 05--08, 2026}{Porto, Portugal}
\acmBooktitle{Interactive Health Conference (IH '26), July 05--08, 2026, Porto, Portugal}
\acmDOI{10.1145/3786579.3803814}
\acmISBN{979-8-4007-2422-0/2026/07}

\begin{document}
\title{Healthcare App Design in Low-Resource Contexts: Challenges, Practices, and Opportunities}

\author{Arka Majhi}
\email{arka.majhi@iitb.ac.in}
\affiliation{%
  \institution{IIT Bombay}
  \city{}
  \country{India}
}
\orcid{https://orcid.org/0000-0002-5057-1878}

\author{Aparajita Mondal}
\email{aparajita.mondal@tuni.fi}
\affiliation{%
  \institution{Tampere University}
  \city{}
  \country{Finland}
}
\orcid{https://orcid.org/0000-0003-4609-2249}

\author{Satish B. Agnihotri}
\email{sbagnihotri@iitb.ac.in}
\affiliation{%
  \institution{IIT Bombay}
  \city{}
  \country{India}
}
\orcid{https://orcid.org/0000-0002-0703-3185}

\renewcommand{\shortauthors}{ Majhi et al.}

\begin{abstract}
Digital health technologies are increasingly used to improve healthcare access and delivery worldwide. However, many healthcare applications are designed for environments with stable infrastructure, high digital literacy, and strong institutional support. These assumptions often do not hold in low-resource contexts where healthcare delivery often depends on community health workers, caregivers, and informal care networks. Designing effective healthcare applications for such environments requires attention to infrastructural constraints, cultural contexts, language diversity, and usability challenges.

This Birds of a Feather session aims to bring together researchers, designers, and practitioners interested in healthcare application design in low-resource contexts. The session will provide an informal forum for discussing challenges encountered in the design and deployment of digital health technologies in underserved settings, sharing field experiences, and identifying opportunities for collaboration within the Interactive Health (IH) community.
\end{abstract}



\maketitle

\section{Introduction and Goal of the BoF}

Digital health technologies such as mobile health applications, telemedicine platforms, and data-driven health systems are increasingly being used to support healthcare delivery, patient engagement, and public health interventions across the world \cite{WHO2021}. Mobile technologies in particular have enabled new forms of healthcare access and communication, especially in regions where traditional healthcare infrastructure may be limited \cite{Kumar2015}.

Despite these advances, many digital health systems are designed with assumptions that reflect high-resource environments with reliable internet connectivity, widespread smartphone adoption, and high levels of digital literacy. In low-resource settings, healthcare delivery rely on community health workers, informal caregivers, and low-cost mobile devices to deliver essential services \cite{Agarwal2015}. These contexts introduce significant challenges for technology design, including low connectivity, device sharing, linguistic diversity, limited technical training, and varying levels of trust in digital systems \cite{Medhi2009}.

Within the Human–Computer Interaction (HCI) community, research areas such as HCI for Development (HCI4D), participatory design, and global health technologies have explored approaches to designing digital systems that are responsive to local needs and contexts \cite{Toyama2011,Dearden2008}. Prior research highlights the importance of community-centered design practices, long-term engagement with stakeholders, and iterative deployment in real-world healthcare environments \cite{Chib2015}. Recent work has also examined how digital tools, mobile applications, and playful technologies can support the training and capacity building of frontline health workers operating in resource-constrained environments \cite{Majhi2021,Majhi2022,Majhi2024HCI_International,Majhi2024CHIPlay,Majhi2024GoodIT,Majhi2026IHSP,Majhi2026IHBoF}.

However, researchers and practitioners working on healthcare technologies for low-resource settings are often distributed across different research communities and disciplinary venues. Opportunities for informal exchange of experiences and lessons learned remain relatively limited within conference programs.

The goal of this Birds of a Feather (BoF) session is to create an informal discussion space where conference attendees interested in healthcare application design for low-resource contexts can meet, exchange experiences, and reflect on design challenges and opportunities.

\section{Target Audience and Relevance to the IH Community}

This BoF session is intended for attendees working in areas related to HCI, digital health, global health innovation, and technology design for underserved communities. Researchers investigating mobile health systems, healthcare technologies for community health workers, participatory design methods, and public health informatics will find the discussion more relevant.

The session will also benefit doctoral students and early-career researchers who are beginning to conduct research in low-resource healthcare environments. These researchers often face complex field conditions that differ significantly from conventional HCI design contexts. Practitioners from non-governmental organizations, public health initiatives, and industry groups developing technologies for healthcare delivery in underserved regions will find value in the discussion. Their practical experiences will enrich conversations about real-world deployment challenges, sustainability, and long-term adoption of digital health interventions.

The topic aligns closely with the goals of the IH conference, which seeks to explore how interactive technologies can support health and well-being across diverse populations and contexts. Designing digital health technologies that function effectively in low-resource settings is essential for ensuring that advances in IH contribute to equitable healthcare access and development.

\section{Session Format and Activities}

The proposed BoF session will run for approximately 60-90 minutes. The session will begin with a short introduction by the organizers outlining the motivation for the discussion and highlighting key challenges in designing healthcare applications for low-resource environments. These challenges include infrastructural constraints, cultural and linguistic diversity, issues of trust and adoption, and the sustainability of digital health interventions.

Participants will then briefly introduce themselves and describe their interest in the topic. This will help participants understand the range of expertise present in the room and facilitate networking among attendees. To encourage interaction and rapid idea exchange, the session will include a short collaborative design reflection activity. Participants will be invited to form small groups and discuss a simple scenario involving the design of a healthcare application for a low-resource setting, such as supporting a community health worker or caregiver in rural or resource-constrained environments. Each group will reflect on the design challenges, contextual constraints, and possible interaction approaches relevant to the scenario. The goal of the activity is not to produce complete solutions but to stimulate discussion about real-world design trade-offs and contextual considerations.

The organizers will facilitate a collective reflection on common themes that emerge during the session, including design strategies, methodological challenges, and opportunities for collaboration. The session will conclude with an open discussion about how researchers and practitioners can continue engaging with this topic beyond the conference.

\section{Organizers and Suitability}

Arka Majhi is a researcher and designer affiliated with the Indian Institute of Technology Bombay whose work focuses on interaction design, digital health communication, and gamification approaches for healthcare engagement. His research explores how design methods and interactive technologies can support healthcare communication and decision-making in contexts where resources, infrastructure, and training opportunities are limited.

Aparajita Mondal is a researcher at Tampere University whose work explores HCI, digital health, and participatory design approaches for socially meaningful technologies. Her research examines how interactive systems can support healthcare engagement and well-being across diverse communities and cultural contexts.

Satish B. Agnihotri is a faculty member at the Indian Institute of Technology Bombay whose work spans engineering systems, healthcare innovation, and technology-enabled public health solutions. His research focuses on the development of practical technological interventions that improve healthcare delivery and health outcomes.

Together, the organizing team brings complementary expertise spanning interaction design, digital health technologies, engineering systems, and global health research. Their collective experience working across academic institutions in India and Europe positions them well to facilitate a productive and inclusive discussion on the challenges and opportunities involved in designing healthcare applications for low-resource contexts.

\section{Promotion and Outreach}

To ensure strong participation, the organizers plan to promote the BoF session through academic mailing lists related to HCI and digital health, announcements on professional networking platforms. During the conference, the organizers will also encourage participation through informal outreach and by sharing information about the session through conference communication channels.

\section{Expected Outcomes}

The BoF session aims to support the development of a stronger community within IH focused on healthcare technology design for low-resource environments. By creating an informal setting for discussion and reflection, the session will enable participants to exchange insights, identify shared challenges, and explore opportunities for collaboration.

In the longer term, the discussions initiated during the session will contribute to future workshops, collaborative research projects, and continued engagement among researchers and practitioners interested in designing inclusive digital health technologies for diverse global contexts.

\subsection{Suggested Locale}

The BoF session will ideally take place in an informal discussion space within the conference venue where participants can sit together in a circular or semi-circular arrangement. Such a setting encourages open dialogue and reduces the formality associated with presentation-based sessions. A whiteboard or flip chart will be useful for capturing key ideas during the discussion.

\bibliographystyle{ACM-Reference-Format}
\bibliography{sample-base}



\end{document}